\documentclass[aps,showpacs,PRB,twocolumn,superscriptaddress]{revtex4-1}
\usepackage{lettrine}
\usepackage{amsmath}
\usepackage{graphicx}
\usepackage{color}
\usepackage{amsfonts}
\usepackage{mathrsfs}
\newcommand{\vdimer}{{\vrule height0.2cm width0.05cm depth0pt}}
\newcommand{\hdimer}{{\hrule height0.05cm width0.2cm depth0pt}}
\newcommand{\verdimers}{\hbox{\vdimer \hskip 0.1cm \vdimer}}
\newcommand{\hordimers}{\hbox{\vbox{\hdimer \vskip 0.1cm \hdimer}}}

\begin{document}

\title{Widely existing mixed phase structure of quantum dimer model on square lattice}
\author{Zheng Yan}
\thanks{These authors contributed equally to this work.}
\affiliation{Department of Physics and State Key Laboratory of Surface Physics, Fudan University, Shanghai 200438, China}
\affiliation{Department of Physics and HKU-UCAS Joint Institute of Theoretical and Computational Physics, The University of Hong Kong, Pokfulam Road, Hong Kong}
\author{Zheng Zhou}
\thanks{These authors contributed equally to this work.}
\affiliation{Department of Physics and State Key Laboratory of Surface Physics, Fudan University, Shanghai 200438, China}
\author{Olav F. Sylju{\aa}sen}
\affiliation{Department of Physics, University of Oslo, P. O. Box 1048 Blindern, N-0316 Oslo, Norway}
\author{Junhao Zhang}
\affiliation{Department of Physics and State Key Laboratory of Surface Physics, Fudan University, Shanghai 200438, China}
\author{Tianzhong Yuan}
\affiliation{Department of Physics and State Key Laboratory of Surface Physics, Fudan University, Shanghai 200438, China}
\author{Jie Lou}
\email{loujie@fudan.edu.cn}
\affiliation{Department of Physics and State Key Laboratory of Surface Physics, Fudan University, Shanghai 200438, China}
\affiliation{Collaborative Innovation Center of Advanced Microstructures, Nanjing 210093, China}
\author{Yan Chen}
\email{yanchen99@fudan.edu.cn}
\affiliation{Department of Physics and State Key Laboratory of Surface Physics, Fudan University, Shanghai 200438, China}
\affiliation{Collaborative Innovation Center of Advanced Microstructures, Nanjing 210093, China}

\begin{abstract}
Quantum dimer model is a low-energy effective model for many magnetic systems (materials) that are candidates for quantum spin liquids. It has strict local constraints which is described by gauge field theory. However, since constraints hinder the application of numerical algorithms, phase diagrams of quantum dimer models are still controversial, even on the square lattice. The core controversy is whether the mixed state exists due to the restriction. In this article, we give strong evidences to solve this dispute. With our newly developed sweeping cluster quantum Monte Carlo method, we studied the phase diagram of large parameter region by introducing the definition of pair correlation function and other supporting evidence to distinguish the mixed phase from the columnar phase with high precision. In particular, we find that the ground state belongs to the mixed phase for a vast parameter region. 
\end{abstract}
\maketitle
\section*{Introduction}
When a physical frustrated system has a particularly large frustrated energy scale, its low energy effective model often contains constraints. Such constraints are common in the formulation of the low-energy description of quantum many-body physics and their feature can usually be captured by lattice gauge theories. As a particularly important example, quantum dimer models~(QDMs) are constraint low energy effective descriptions of certain quantum spin systems~\cite{RK1988,Misguich2003,Poilblanc2010}. QDMs were first introduced by Rokhsar and Kivelson~(RK) to study the physics of the short-range resonating valence bond~(RVB) state in potential relation to high-$T_c$ superconductor~\cite{Anderson1987,Fazekas1974,Kivelson1987,ZhengZhou2019}. QDMs provide particularly simple examples to realize topological phases, such as a two-dimensional gapped phase with $\mathbb{Z}_2$ topological order~\cite{Moessner2001,yan2020triangular}, and a three-dimensional Coulomb phase described by an emergent $U(1)$ symmetry~\cite{Hermele2004,Huse2003,Patil2021PRB}. In addition to the spin liquids, QDMs are also an important carrier of incommensurate phases~\cite{XF1,XF2,M3,cantor_deconfinement}. Recently, a QDM for the metallic state of the hole-doped cuprates was also proposed to describe the mysterious pseudogap state at low hole density~\cite{Sachdev2015}.

However, the strong geometrical constraint present in the QDMs hampers the application of an numerical algorithms. It is thus imperative to find accurate numerical algorithms that can treat such systems efficiently, without which research and progress in understanding constraint systems will be delayed heavily. As a result, the phase diagrams of QDMs are still controversial, even on the square lattice. The QDM Hamiltonian on the square lattice can be written as
\begin{equation}
    H=-\sum_{\rm plaq}\left(\vphantom{\sum}|\verdimers\rangle\langle\hordimers|+\rm{H.c.}\right)
       +V\sum_{\rm plaq}\left(\vphantom{\sum}|\verdimers\rangle\langle\verdimers|+
                                             |\hordimers\rangle\langle\hordimers|\right)
\label{Hamiltonian}
\end{equation}
where the summations are taken over all elementary plaquettes of the lattice. The kinetic term describes the resonance between the two dimerizations of a plaquette, while the potential term counts the plaquettes on which resides two parallel dimers. In addition, strong geometric constraints are imposed on the Hilbert space which requires every site on the lattice to be covered by one and only one dimer.

\begin{figure}[htp]
\includegraphics[width=\linewidth]{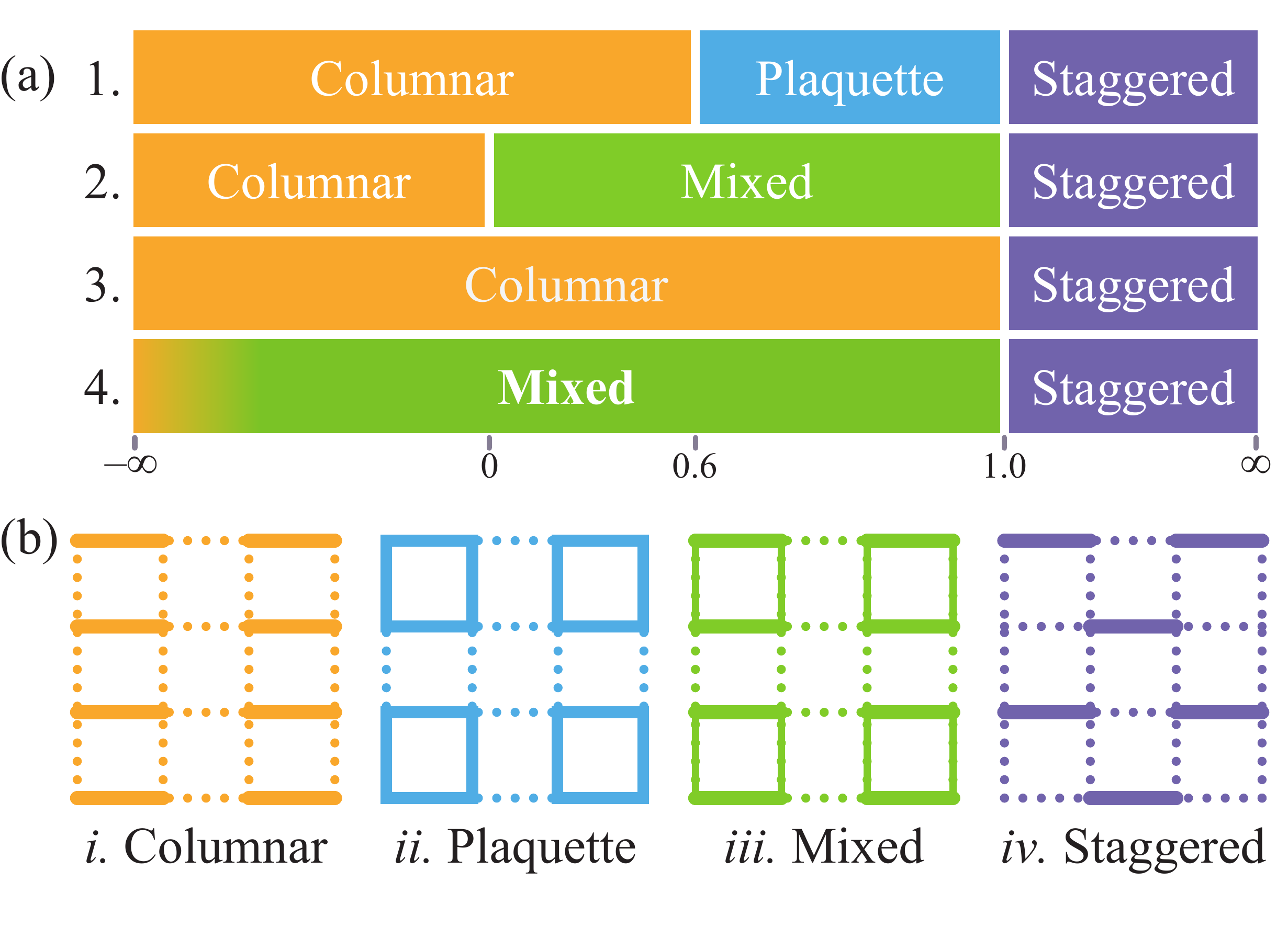}
\caption{(a) Possible phase diagrams of QDM on square lattice. 1 -- Ref.~\cite{OFS2005,OFS2006}, 2 -- Ref.~\cite{Ralko2008}, 3 -- Ref.~\cite{Banerjee2014,Banerjee2016}, 4 -- this article; (b) Schematic diagram of the four possible phases.}
\label{Fig1}
\end{figure}

On the phase diagram exists one particularly important point, namely, the Rokhsar-Kivelson (RK) point~($V=1$), at which the ground state of Hamiltonian is exactly solvable as a pure RVB state:
\begin{equation}
    |\textrm{GS}\rangle=\sum_{C}A_C|C\rangle
\label{RVB}
\end{equation}
where $C$ is a dimer covering and $A_C = A_{C'}$ for $C$ and $C'$ in same winding sector~\cite{QDMbook}. However, the model, Eq.~(\ref{Hamiltonian}), cannot be solved exactly at other parameters, and there are still disputes about its phase diagram, as discussed below and illustrated in Fig.~\ref{Fig1}(a).

When $V>1$, the staggered phase (Fig. \ref{Fig1}(b)iv) with no face-to-face (in the same plaquette) dimers is favoured, whereas on the other side of the RK point, the phase diagram is less clear. In the limit $V\to-\infty$, the Hamiltonian strongly favors configurations with as many parallel dimers as possible, known as the columnar phase (Fig. \ref{Fig1}(b)i). However, when increasing $V$, the quantum effect of resonance brought by the kinetic term becomes more prominent, which brings more possibilities to the phase diagram. Candidate phases are, plaquette phase which breaks translation symmetry along both axes and respects the four-fold rotation symmetry (Fig. \ref{Fig1}(b)ii), and mixed phase which breaks the translation symmetry along two axes as well as the rotation symmetry (Fig. \ref{Fig1}(b)iii). Mixed phase configurations look similar to the plaquette phase, except that the strengths of the bonds along $x$ and $y$ directions of the same plaquette are different. Its nature is still controversial. Whether it is a mixture of plaquette phase and columnar phase or an independent quantum state is disagreed upon.

How these candidate phases enter the phase diagram remains a disputed issue. A projection Monte Carlo study has found a plaquette phase adjacent to the RK point and a plaquette-columnar phase transition at $V\sim0.6$~\cite{Leung1996,OFS2005,OFS2005walk,OFS2006} (Fig. \ref{Fig1}(a)1). However, through an exact diagonalization and Green's function Monte Carlo study, some has concluded that it is a mixed phase instead of a plaquette phase that resides in vicinity of the RK point ~\cite{Ralko2008} (Fig. \ref{Fig1}(a)2). There were also arguments that columnar state extends all the way up to the RK point~\cite{Sachdev1989} (Fig. \ref{Fig1}(a)3), supported by Metropolis Monte Carlo simulations on height model equivalents of the square lattice QDM~\cite{Banerjee2014,Banerjee2016} and the frustrated transverse field Ising model (TFIM) which is equal to a parameter point ($V=0$) of QDM~\cite{mila2012}.




In this paper, we are committed to solving the phase diagram dispute and giving a result that reconciles all contradictions. Using our newly developed numerically exact method -- sweeping cluster algorithm~\cite{ZY2019,yan2020improved}, we calculate the phase diagram of the square lattice QDM and find the 4th one of Fig.~\ref{Fig1}(a).

\begin{figure}[b]
    \includegraphics[width=\linewidth]{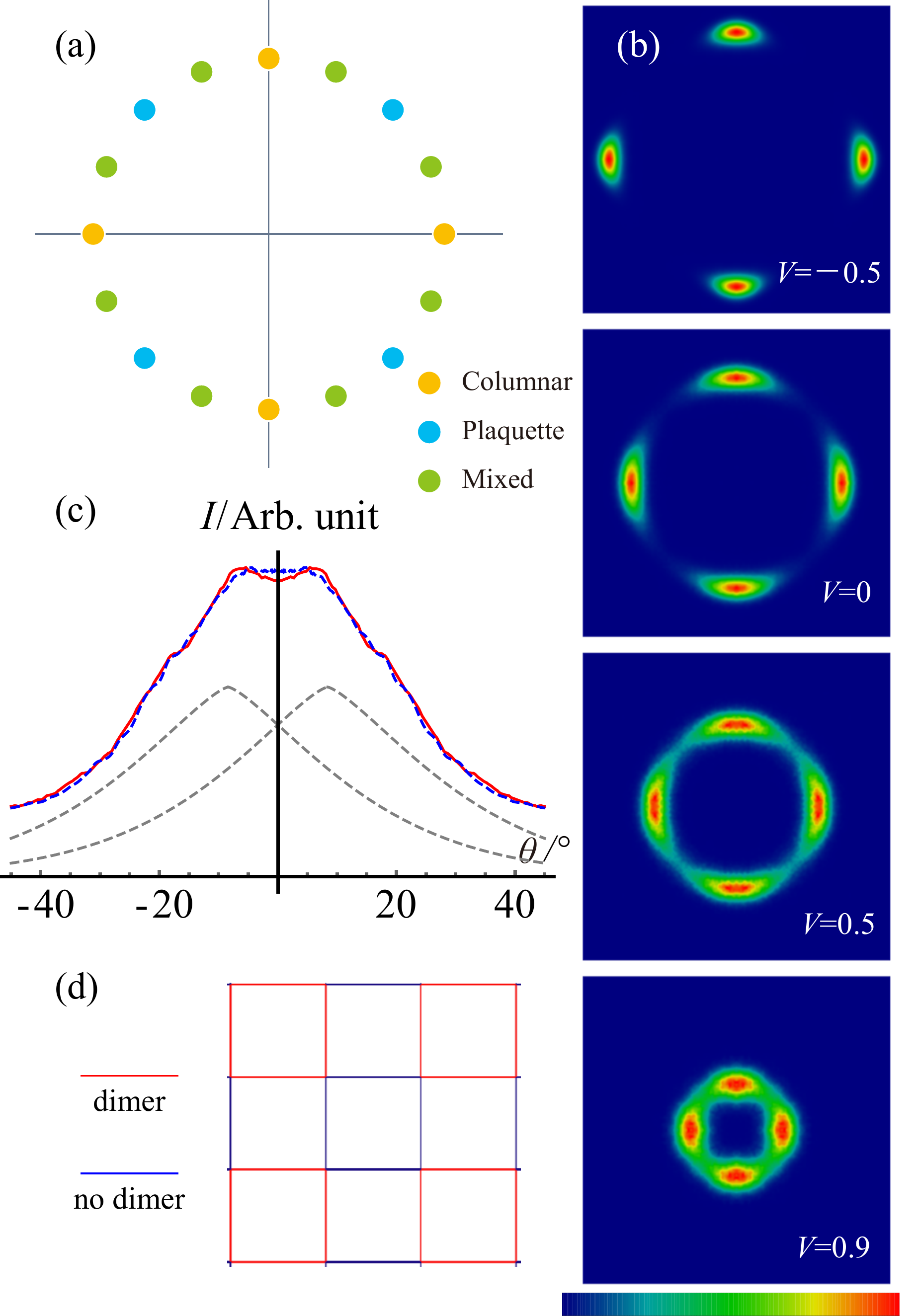}
    \caption{(a) An illustration of VBS order parameter distributions of the various candidate phases. (b)~VBS order parameter distributions of different $V$, about 80,000 data points are taken for each histogram. $T=0.01$, $L=32$ is taken. (c) The angular distribution of the VBS order parameter measured at $V=0.5$ obtained through integrating out the radial distribution (blue dashed lines) and cut through a fixed radius (red solid lines), and the double peak fit (grey dashed lines). (d)~The average dimer occupation near the center of an $L=64$ lattice for $V=0.5$. Red/blue color represents that the dimer occupation is larger/smaller than $1/4$ (the average number when no long range order exists), indicating the tendency to find one/no dimer at that location. We choose the parameters as $V=0.5$, $T=0.01$. The details about the peak fitting is written in the Appendix A of Supplemental Material~\cite{SM}.}
    \label{Fig2}
\end{figure}

\section*{Mixed Phase and Phase Diagram}
There have been strong evidences supporting that the plaquette phase do not exist on square lattice~\cite{Ralko2008}. So the first question is whether there exists a mixed phase or there is only the columnar phase, and whether the mixed phase is a multiphase mixture or an individual state independent of the columnar or plaquette phase. Selecting $V=0.5$, we carefully studied the ground state. Different states can be distinguished by different distributions of VBS order parameter, defined as~\cite{Sachdev1989}
\begin{equation}
    \begin{split}
        \Psi_{\mathrm{col}}=\frac{1}{L^2}\sum_{\bf r}&\left\{
        (-1)^{r_x} [n({\bf r}+\frac{\hat{\bf x}}{2})-n({\bf r}-\frac{\hat{\bf x}}{2})]+\right.\\
                            &\left.i(-1)^{r_y}[n({\bf r}+\frac{\hat{\bf y}}{2})-n({\bf r}-\frac{\hat{\bf y}}{2})]\right\},
        \label{ficol}
    \end{split}
\end{equation}
where ${\hat{\bf x}}$ and ${\hat{\bf y}}$ are unit vectors and $L$ is the linear system size. The dimer number operator $n({\bf r}+{\bf e}/2)$ is 1 if a dimer resides on the link connecting ${\bf r}$ and its nearest neighbor at ${\bf r}+{\bf e}$, and zero otherwise. The peaks of its histogram distinguishes three different candidate phases. As shown by Fig.~\ref{Fig1}(a), the yellow points represent the columnar state with 4-fold degeneracy; the blue ones represent the plaquette state with 4-fold degeneracy; the green ones between blue and yellow points indicate mixed state which has 8-fold degeneracy. It is worth noting that the green points are not necessarily the exact middle point of yellow and blue. It can move in the region depending on the degree of mixing. A mixture of columnar phase and plaquette phase would be indicated by 8 peaks at both yellow and black points.

The VBS order parameter distribution Fig.~\ref{Fig2}(b) peaks at the location of the yellow points in Fig.~\ref{Fig2}(a). This has been taken as the main evidence for the columnar state in Ref.~\cite{Banerjee2014,mila2012}. However, when we scrutinize one of the peaks and plot its distribution as a function of the order parameter angle $\theta$, we observe that the distribution has a flat maximum which can equally well be interpreted as a combination of two peaks centered on $\pm \theta_0$ to the sides of the yellow point, as we have fitted in Fig.~\ref{Fig2}(c). Thus it is possible to interpret the flat maxima at the ``columnar'' points of the order parameter distributions as two mixed-phase peaks instead. This scenario become clearer if we fix a certain radius, whose angular distribution looks more like two distinct peaks (red line, Fig.~\ref{Fig2}(c)). 
%
\begin{figure}[hpt]
    \begin{minipage}{\linewidth}
        \includegraphics[width=\linewidth]{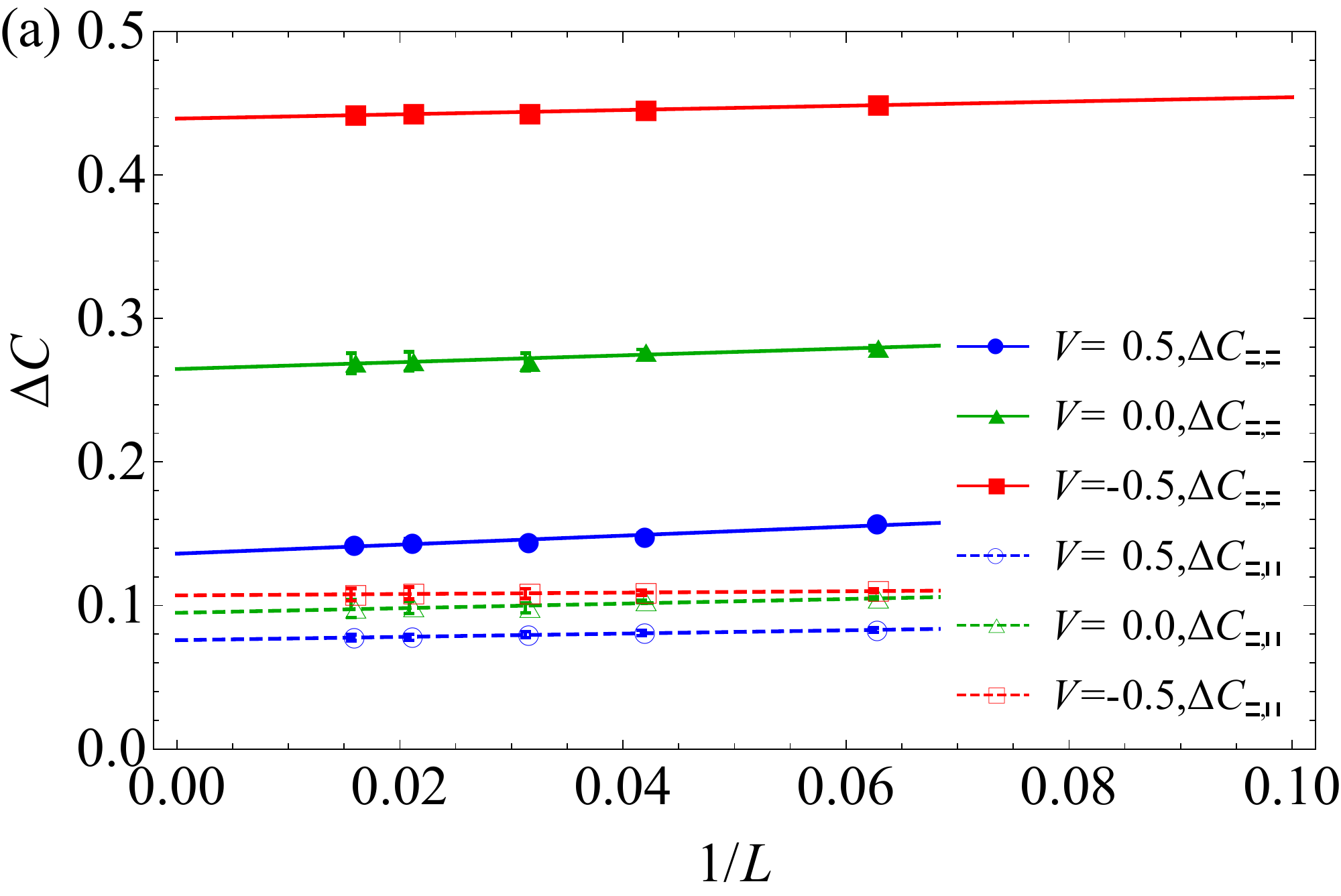}
    \end{minipage}
    \begin{minipage}{\linewidth}
        \includegraphics[width=\linewidth]{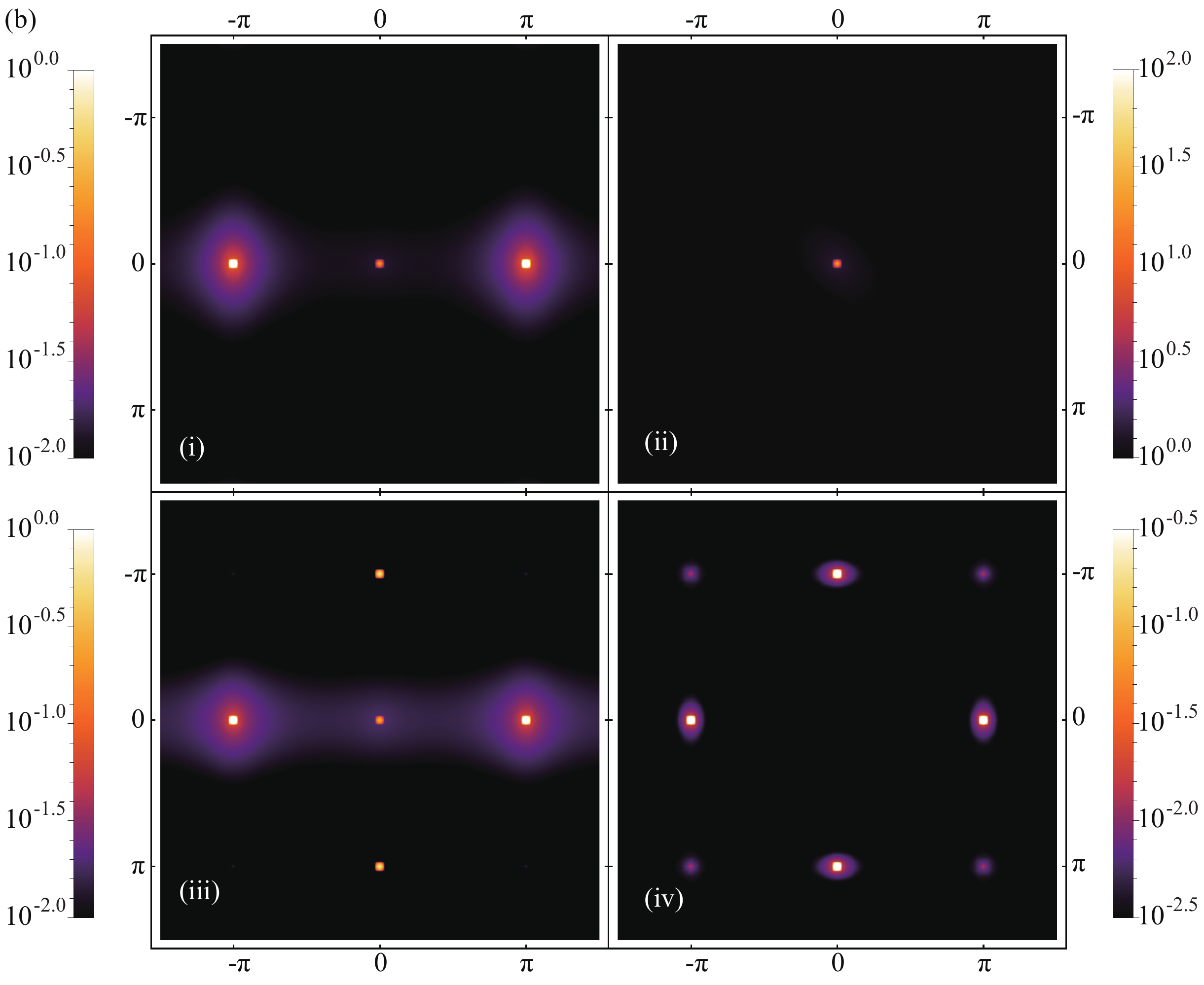}
    \end{minipage}
    \begin{minipage}{\linewidth}
        \includegraphics[width=\linewidth]{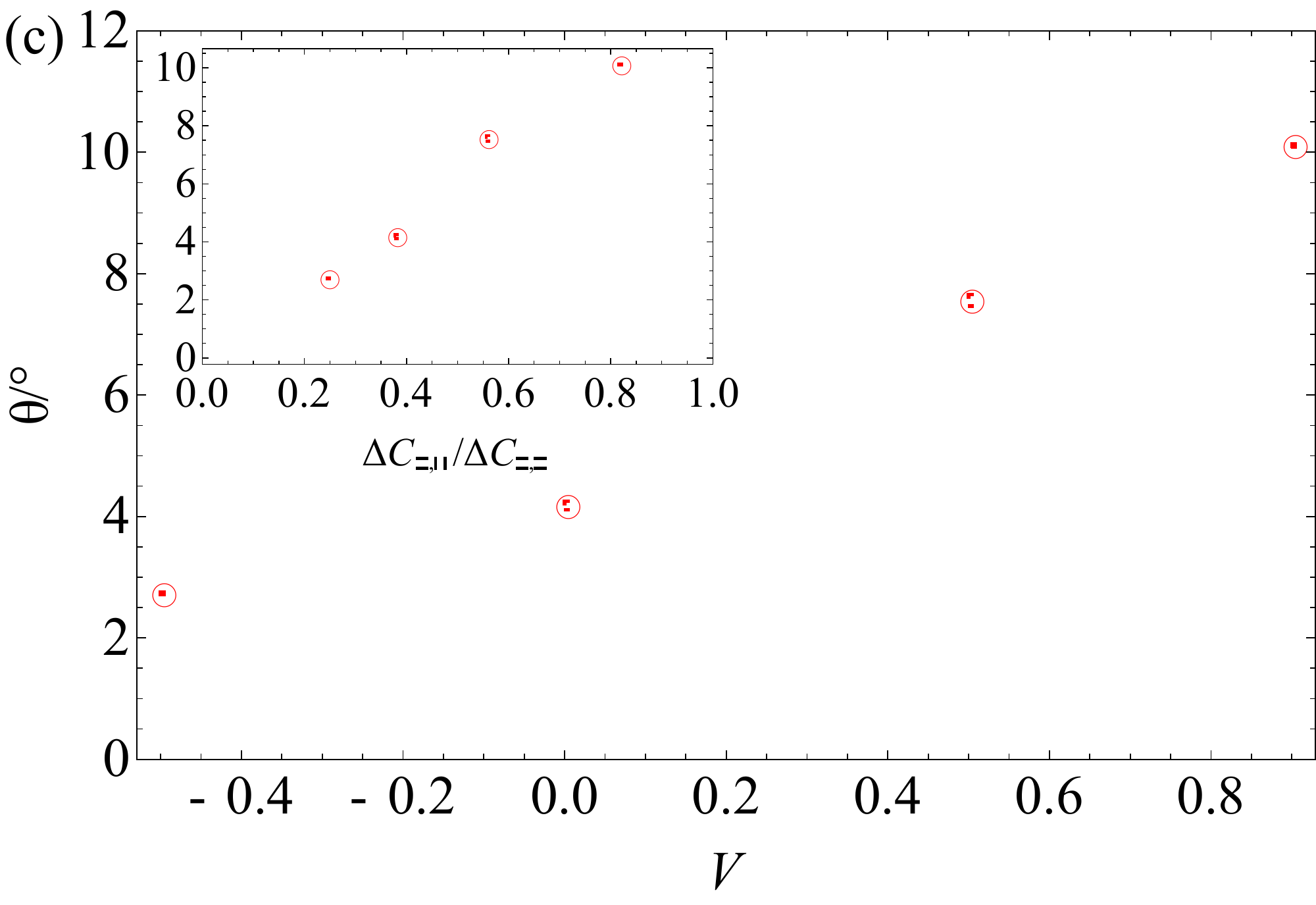}
    \end{minipage}
    \caption{(a)~The difference of longest distance pair dimer correlations $\Delta C_{\hordimers,\hordimers}$(red) and $\Delta C_{\hordimers,\verdimers}$(blue). (b)The structure factor of different dimer correlation functions under size $L=64$, temperature $T=0.01$ and parameter $V=0$. (i) Single dimer correlation function $C_{-,-}$. (ii) Single dimer correlation function $C_{-,|}$. (iii) Pair correlation function $C_{\hordimers,\hordimers}$. (iv)Pair correlation function $C_{\hordimers,\verdimers}$. (c)~The relationship of the distance of two mixed state peaks $\theta_0$ and V under certain size $L=32$ and temperature $T=0.01$. Inset: The relation for various $V$ ($V=-0.5,0,0.5,0.9$) between distance of two mixed state peaks $\theta_0$ and comparation with $\Delta C_{\hordimers,\verdimers}/\Delta C_{\hordimers,\hordimers}$, $L=32$ and $T=0.01$.}
    \label{Fig3}
\end{figure}

It is hard to distinguish different states by average configurations directly since the features are lost when averaging all the degenerate states. To remove the degeneracy we have to act a projection operator on the samples of Monte Carlo data to remove the unwanted configurations. Specifically we project an operator $|\hordimers\rangle_n\langle \hordimers|_n$ on one specific plaquette (labelled $n$) on the square lattice, it is non-zero only when the plaquette contains two parallel horizontal dimers. As shown in Fig. \ref{Fig2}(d), the averaged projected configuration shows clear evidence against the columnar phase.

To seek stronger evidence, we measure dimer pair correlation functions. We define the pair dimer operator on plaquette at position $\mathbf{r}$ as
\begin{equation}
\begin{aligned}
    D_{\hordimers,\mathbf{r}}&=|\hordimers\rangle_\mathbf{r}\langle\hordimers|_\mathbf{r}, \\
    D_{\verdimers,\mathbf{r}}&=|\verdimers\rangle_\mathbf{r}\langle\verdimers|_\mathbf{r}.
\end{aligned}
\end{equation}
and the pair correlation function $C_{ij,\mathbf{r}-\mathbf{r}'}$ between $D_{i,\mathbf{r}}$ and $D_{j,\mathbf{r}'}$ as:
\begin{equation}
    C_{ij,\mathbf{r}-\mathbf{r}'}=\frac{\langle D_{i\mathbf{r}}D_{j,\mathbf{r}'}\rangle-\langle D_{i,\mathbf{r}}\rangle\langle D_{j,\mathbf{r}'}\rangle}{\langle D_{i,\mathbf{r}'}\rangle-\langle D_{i,\mathbf{r}'}\rangle^2}
\end{equation}
where $i,j=\hordimers,\verdimers$ and $\mathbf{r}-\mathbf{r}'$ is the position difference. We investigate the difference between the correlations of the two largest distance, \textit{i.e.}, with $\mathbf{r}-\mathbf{r}'=\frac{L}{2}\hat{\mathbf{x}}+\frac{L}{2}\hat{\mathbf{y}}$ and $\frac{L}{2}\hat{\mathbf{x}}+\left(\frac{L}{2}+1\right)\hat{\mathbf{y}}$
\begin{equation}
    \left.\begin{aligned}
        &\Delta C_{\hordimers,\hordimers}=C_{\hordimers,\hordimers,(L/2)\hat{\mathbf{x}}+(L/2)\hat{\mathbf{y}}}-C_{\hordimers,\hordimers,(L/2)\hat{\mathbf{x}}+(L/2+1)\hat{\mathbf{y}}}\\
        &\Delta C_{\hordimers,\verdimers}=C_{\hordimers,\verdimers,(L/2)\hat{\mathbf{x}}+(L/2)\hat{\mathbf{y}}}-C_{\hordimers,\verdimers,(L/2)\hat{\mathbf{x}}+(L/2+1)\hat{\mathbf{y}}}
    \end{aligned}\right\}
\end{equation}
At $V=0.5$ and $T=0.01$, we plot those for different system sizes in Fig.~\ref{Fig3}.

In a columnar phase, $\Delta C_{\hordimers,\hordimers}$ should extrapolate to a finite value as $L \to \infty$, while $\Delta C_{\hordimers,\verdimers}$ scales to 0. For a plaquette phase $\Delta C_{\hordimers,\hordimers}=\Delta C_{\hordimers,\verdimers}$, while a mixed phase is characterized by finite but different values of $\Delta C_{\hordimers,\hordimers}$ and $\Delta C_{\hordimers,\verdimers}$ in the same limit. As shown in Fig.~\ref{Fig3}(a), our results taken from system sizes up to $L=64$ indicate a mixed phase. We can conclude here for the $V=0.5, 0, -0.5$ results that there are substantial mixed phase correlations.

It is worth noting that such distinction between columnar phase and mixed phase can not be seen in the single dimer correlation function $C_{-,-}$ and $C_{-,|}$. To illustrate that, we measured various structure factor, \textit{i.e.}, the Fourier transformation of the dimer correlations $C_{-,-}$ and $C_{-,|}$ as well as the pair correlation functions $C_{\hordimers,\hordimers}$ and $C_{\hordimers,\verdimers}$ into the momentum space. We choose $V=0$, at which projection algorithm loses its effectiveness and the finite size effect becomes non-significant. The pair structure factor exhibits two additional peaks at $(0,\pi)$ compared with the columnar phase, clearly confirming that the ground state is the mixed phase, while the single dimer structure factor shows no such different.


Both a finite $\Delta C_{\hordimers,\verdimers}/\Delta C_{\hordimers,\hordimers}$ and the splitting of peaks $\theta_0$ of the complex order parameter are features of mixed state. We find these two quantum positively related under various $V$ (inset of Fig.~\ref{Fig3}(c)), which has further confirmed out our starting point, i.\,e. the peak for columnar state is made up by two peaks for columnar state. The mixed phase here is not a multiphase mixture but a single state with both rotational symmetry and translational symmetry broken. In addition, both $\Delta C_{\hordimers,\verdimers}/\Delta C_{\hordimers,\hordimers}$ and $\theta_0$ do not decay rapidly when $V<0$ from this figure. This means that the mixed state may extend to an area of small $V$. Since the distributions of VBS order parameter can only be obtained in finite sizes, further study is needed.

\begin{figure}[htp]
    \centering
    \includegraphics[width=\linewidth]{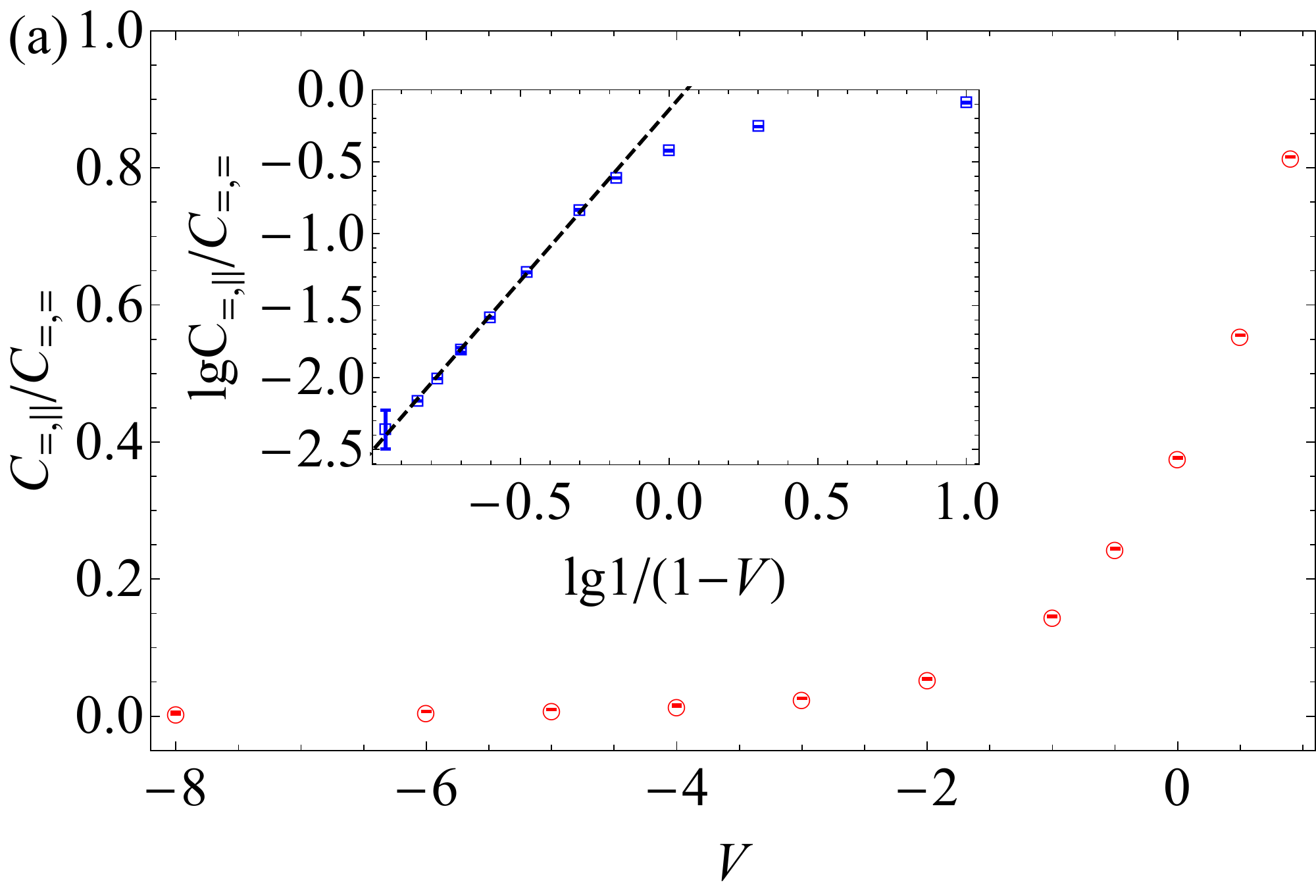}
    \caption{The relationship of $\Delta C_{\hordimers,\verdimers}/\Delta C_{\hordimers,\hordimers}$ and $V$. Inset: The relationship of $\ln\Delta C_{\hordimers,\verdimers}/\Delta C_{\hordimers,\hordimers}$ and $\ln 1/(1-V)$ is power law. All the data in (a) are extrapolated from the finite size. The original data is shown in the Appendix C of Supplemental Material~\cite{SM}.}
    \label{Fig4}
\end{figure}

Having confirmed the existance of mixed phase, our second question concerns the boundary of the mixed state. We aim to make clear the entire phase diagram. Based on field theory analysis combined with exact diagonalization method, recent studies have shown that there should be no phase transition points from the classical limit~($V=-\infty$) to the RK point~\cite{Banerjee2014,Banerjee2016}. A similar trend can also be seen in Fig.~\ref{Fig3}(c), the distance between two peaks of the mixed state also tends to a non-zero finite value though this is a result of finite size. In addition, we have also measured difference of pair correlation function at $V=-0.5$ and got the similar result that there is still mixed phase, as Fig.~\ref{Fig3}(a).

However, the scenario is still unclear when $V$ is far less than $-0.5$. When $V$ is less than $-1$, for convenience, we take the energy shift $C=-v$ in sweeping cluster method, such that the algorithm returns to the ``pair update'' which flip only face-to-face dimers. ``Pair update'' works well when applied far from RK point on square lattice (See Appendix B of Supplemental Material~\cite{SM}). This scheme effectively switches between the columnar state and the plaquette state. We set the initial state of QMC as the columnar state and performed ``pair update'' to see whether the simulation equilibrates in the mixed phase. 
Then we have done a finite $V$ scaling as Fig.~\ref{Fig4}(a). The $\Delta C_{\hordimers,\verdimers}$ which indicates the characteristics of the mixed state, is always non-zero and show power law dependence on $1/(1-V)$. It seems that as long as there is a quantum fluctuation term, the system is always in a mixed state, even if it is very small. From these proofs, we could have a clear cognition about the phase diagram of QDM on square lattice. Close to classical limit, columnar phase is the ground state. When we add kinetic term into the Hamiltonian, it becomes a mixed state for a vast parameter region. After $V>1$~(RK point), the conclusion remains that the system enters a staggered state.

\section*{Discussion and Outlook}
Although it seems that the structure of mixed phase extends to the classical limit ($V=-\infty$), this trend is not exactly true. 
In the classic limit, the ground state is a pure columnar state with gapped exitations. The different degeneracy between two phases, \textit{i.e.}, the columnar phase is 4-fold degenerate and the mixed phase is 8-fold degenerate, forbids a smooth crossover between the two phases and dictates the existence of a second order phase transition point between them with gap closed. At the same time, there must be a vast mixed phase according to the numerical results. It seems that the phase transition point is far from RK point. 
The mixed structure seems to have made a small kinetic energy correction while ensuring the optimal potential energy of the columnar-like main ingredient, allowing the most plaquettes to resonate to achieve the overall optimization of kinetic energy and potential energy.

An additional clue of the widely existence of the mixed phase comes from the ($2+1$)-dimensional $U(1)$ quantum link model~\cite{DH1990,PO1990,SC1997,chakravarty2002theory,shannon2004cyclic} closely related to the square lattice quantum dimer model, in which there exist in particular two distinct confining phases~(analogous to columnar and plaquette phases in the quantum dimer model) with different discrete symmetry breaking patterns, separated by a weak first-order phase transition that mimics several features of deconfined quantum critical points~\cite{AV2004,TS2004a,banerjee20132,tschirsich2019phase}. This implies that there should also be a similar ordered phase other than the columnar in QDM.

The existence of the mixed state in the square lattice dimer model has been controversial for a long time. In recent articles, researchers have denied the possibility of mixed phase, and proposed that only columnar states exist in its phase diagrams. This article provides strong numerical evidences to prove its existence through detailed analysis of the histogram of the order parameter and the anisotropy of the pair correlation function. Furthermore, we find that it exists in a wide range even far from the RK point.

Further, we will study the physics of the constrained system at a finite temperature\cite{Alet2005PRL,Alet2006PRE} and the characteristics of the restricted system under dissipation\cite{ZiCai2013PRL,ZiCai2013PRA,ZY2018}.

\section*{Acknowledgements}
We wish to thank Andereas Lauchili, Xue-Feng Zhang, Wei Li, Zi Yang Meng, Yang Qi and Yuan Wan for fruitful discussions. This work is supported by the National Key Research and Development Program of China (Grants Nos. 2017YFA0304204 and 2016YFA0300504), the National Natural Science Foundation of China Grant No. 11625416, and the Shanghai Municipal Government (Grants Nos. 19XD1400700 and 19JC1412702). J.~L. thanks the support of National Natural Science Foundation of China Grant No. 11304041. Z.~Z. acknowledges for the support by the CURE (Hui-Chun Chin and Tsung-Dao Lee Chinese Undergraduate Research Endowment) (19925) and National University Student Innovation Program (19925). Z.~Y. acknowledges the support provided by the Kavli Institute for Theoretical Sciences (KITS) while in Beijing.

%

%

\setcounter{page}{1}
\setcounter{equation}{0}
\setcounter{figure}{0}
\renewcommand{\theequation}{S\arabic{equation}}
\renewcommand{\thefigure}{S\arabic{figure}}

\newpage
\begin{widetext}
\section{Supplemental Material}


\centerline{}

\title{Supplemental Material for "Widely existing mixed phase structure of quantum dimer model on square lattice"}
\author{Zheng Yan}
\thanks{These authors contributed equally to this work.}
\affiliation{Department of Physics and State Key Laboratory of Surface Physics, Fudan University, Shanghai 200438, China}
\affiliation{Department of Physics, The University of Hong Kong, Hong Kong, China}
\author{Zheng Zhou}
\thanks{These authors contributed equally to this work.}
\affiliation{Department of Physics and State Key Laboratory of Surface Physics, Fudan University, Shanghai 200438, China}
\author{Olav F. Sylju{\aa}sen}
\affiliation{Department of Physics, University of Oslo, P. O. Box 1048 Blindern, N-0316 Oslo, Norway}
\author{Junhao Zhang}
\affiliation{Department of Physics and State Key Laboratory of Surface Physics, Fudan University, Shanghai 200438, China}
\author{Tianzhong Yuan}
\affiliation{Department of Physics and State Key Laboratory of Surface Physics, Fudan University, Shanghai 200438, China}
\author{Jie Lou}
\email{loujie@fudan.edu.cn}
\affiliation{Department of Physics and State Key Laboratory of Surface Physics, Fudan University, Shanghai 200438, China}
\affiliation{Collaborative Innovation Center of Advanced Microstructures, Nanjing 210093, China}
\author{Yan Chen}
\email{yanchen99@fudan.edu.cn}
\affiliation{Department of Physics and State Key Laboratory of Surface Physics, Fudan University, Shanghai 200438, China}
\affiliation{Collaborative Innovation Center of Advanced Microstructures, Nanjing 210093, China}

\maketitle

\subsection*{APPENDIX A: Peak fitting of the VBS order distribution}
\begin{figure*}[htb]
    \includegraphics[clip,width=17cm]{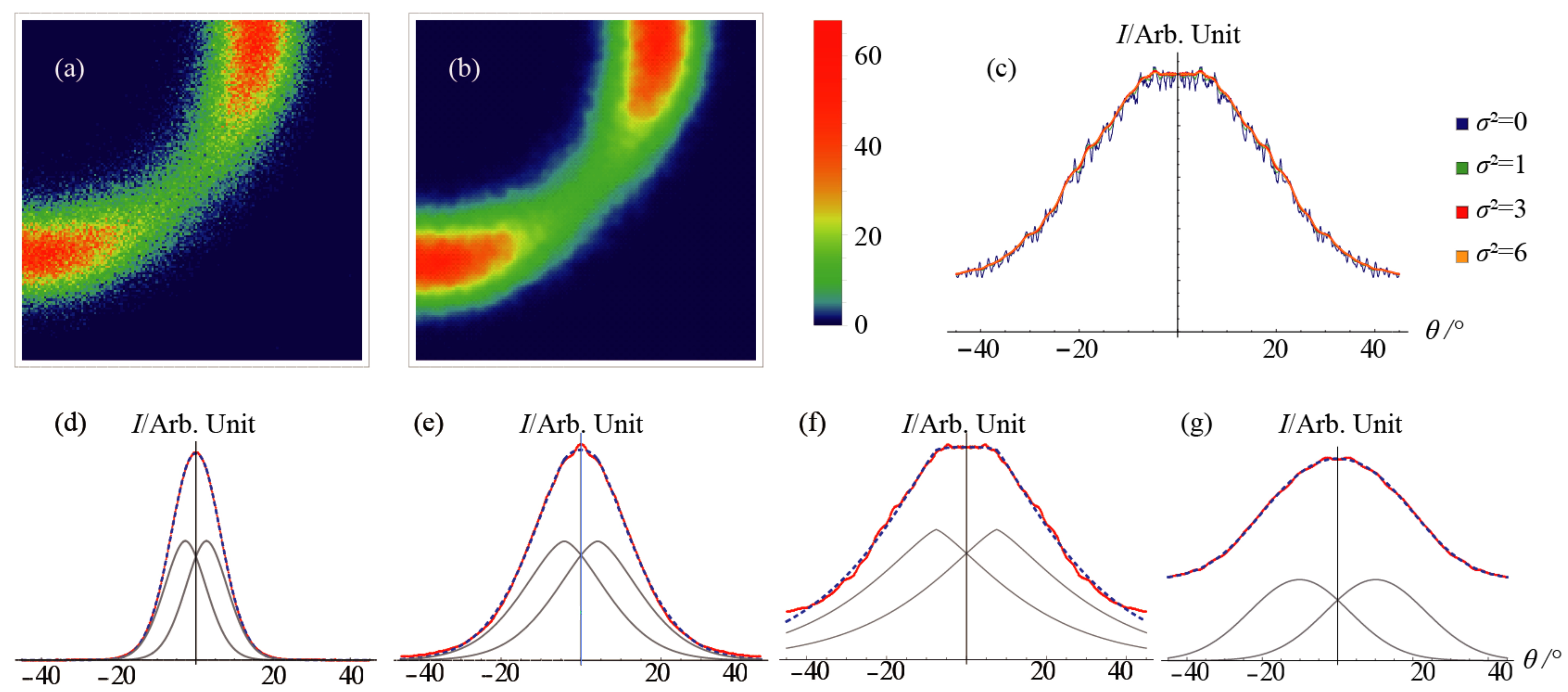}
    \caption{(color online). (a)(b)~The 2D distribution of VBS order parameter (a)~before and (b)~after an image smoothing; (c)~The angular distribution for various smoothing radius $\sigma$; (d)--(g)~The peak fittings of the angular distribution for (d)~$V=-0.5$, (e)~$V=0$, (f)~$V=0.5$ and (g)~$V=0.9$. }
    \label{Fig5}
\end{figure*}

\begin{figure*}[htb]
    \includegraphics[clip,width=8.6cm]{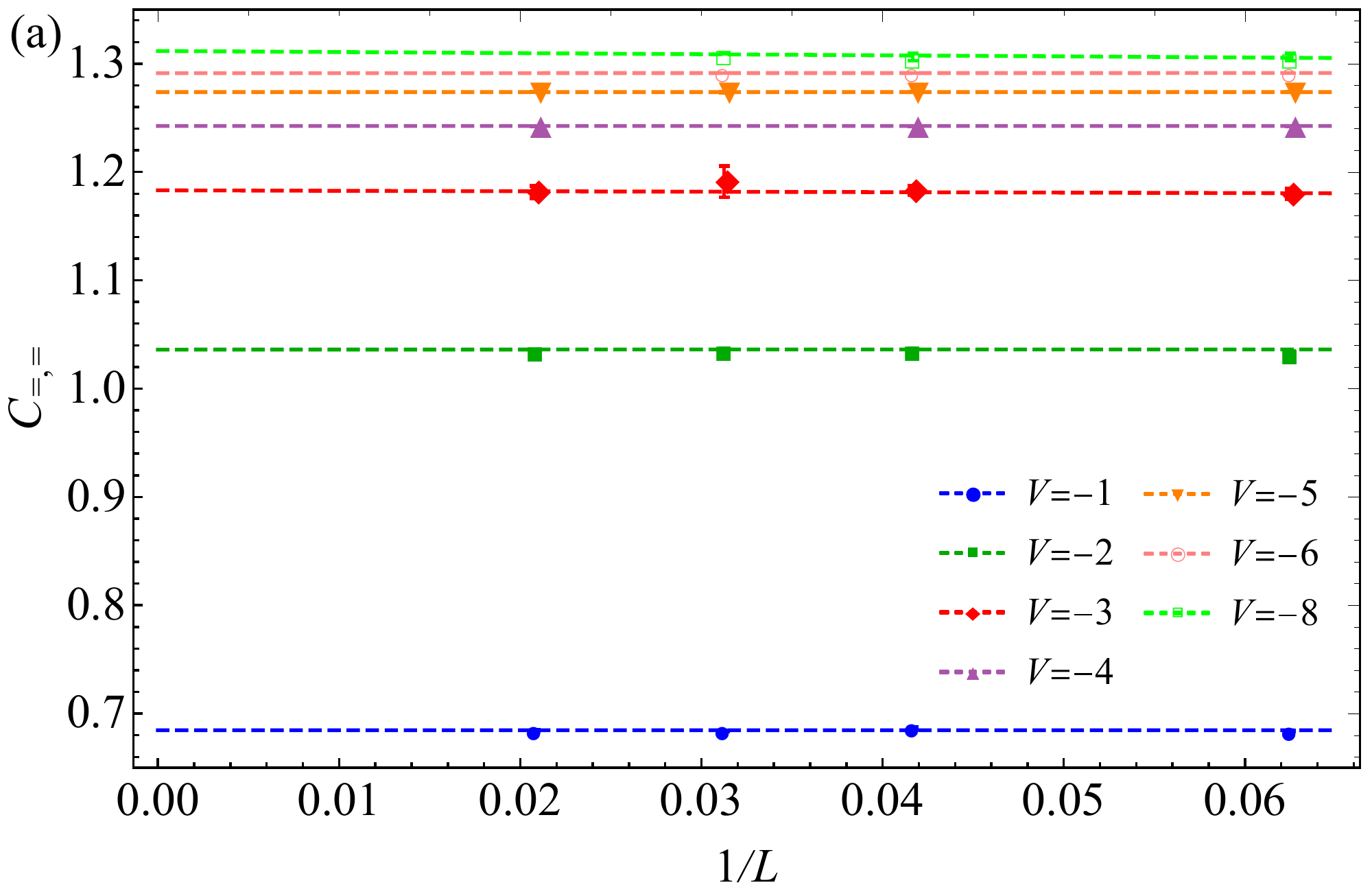}
    \includegraphics[clip,width=8.6cm]{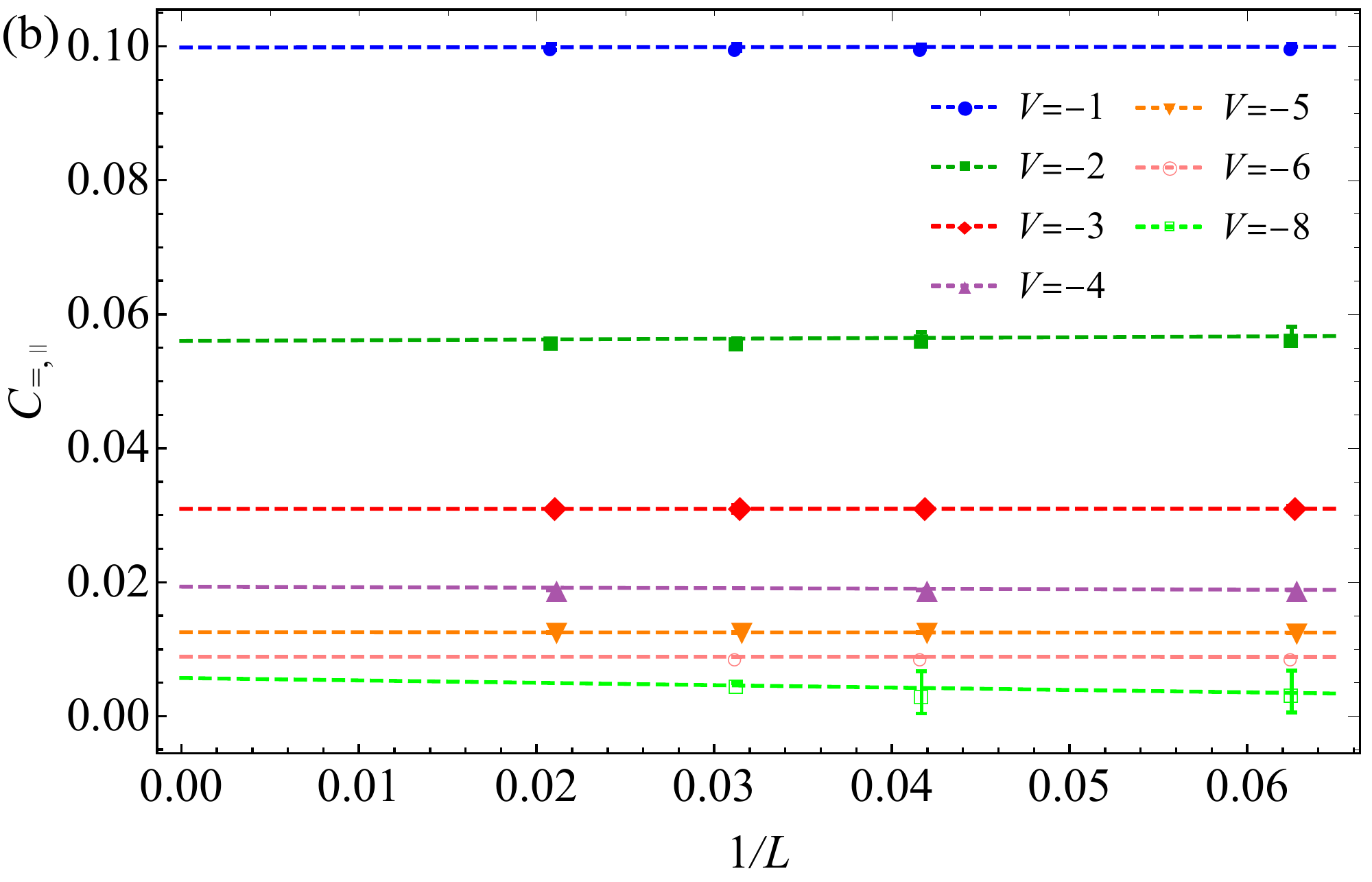}
    \caption{(color online). (a)~The relationship of $\Delta C_{\hordimers,\hordimers}$ and size $1/L$ for different $V$ at T=0.01. (b)~The relationship of $\Delta C_{\hordimers,\verdimers}$ and size $1/L$ for different $V$ at T=0.01.}
    \label{Fig6}
\end{figure*}

This appendix will give a detailed description on the approach of distinguishing the peaks from the 2D distribution of VBS order parameter, i.e., Fig\ref{Fig5}(a)(b) of appendix(and Fig.2(b) of article).

The points we obtained are initially discretely distrubuted on grid points $(4m,4n)$. To obtain a continuous angular distribution, certain kinds of loyal smoothing and continuation are neccesary.

We first carry out an image smoothing with radius $\sigma$ to reduce the noise:
\begin{equation}
    \bar{N}_{mn}=\mathcal{C}\sum_{m'n'}N_{m'n'}\exp\left(\frac{(m'-m)^2+(n'-n)^2}{2\sigma^2}\right).
\end{equation}
We choose $\sigma=\sqrt{3}$ because smoothing with this radius can make the image smooth enough without flattening the peaks, shown in Fig.~\ref{Fig5}(b). The angular distributions for various $\sigma$ are shown in Fig.~\ref{Fig5}(c). We can see that $\sigma^2=3$ is the most satisfying.

The continuous 2D distribution $\varphi(x,y)$ is obtained through a linear interpolation from the discrete distribution at the grid points $\bar{N}_{mn}$. After intergrating out the radial part, we get the continuous angular distribution

\begin{equation}
    \varphi(\theta)=\int r\mathrm{d}r\varphi(r,\theta)
\end{equation}

The peaks from the angular distribution don't follow the Gauss $\exp(-Ax^2)$ type very well, so we use $\exp(-Ax^{\alpha})$ to fit our peaks, where $\alpha$ is a tunable parameter. We carry out a four-parameter fitting
\begin{equation}
    \varphi(\theta)=C\big(\mathrm{e}^{-A(\theta-\theta_0)^\alpha}+\mathrm{e}^{-A(\theta+\theta_0)^\alpha}\big)
\end{equation}
If $\theta_0=0$, then the distribution is single-peaked, and the system is in columnar state. If $\theta_0\neq 0$, the system is in mixed state. The fitting of $V=0.9$ data will have to add a constant term because of the influence of the orderless phase at RK point.

We carried out fittings for $V=-0.5,0,0.5$ and $0.9$, shown in Fig.~\ref{Fig5}. We find that the ratio of distance of peaks to $\Delta C_{\hordimers,\verdimers}/\Delta C_{\hordimers,\hordimers}$ is proportional.

\begin{table}
    \begin{tabular}{ccc|crc|crc}
        \hline
        &$r$&&&\multicolumn{1}{c}{$C=-V$}&&&\multicolumn{1}{c}{$C=1$}&\\
        \hline
        &0 &&&$1.000000(000) $&&&$ 1.000000(000)$&\\
        &1 &&&$-0.333333(000)$&&&$-0.333333(000)$&\\
        &2 &&&$0.380892(136) $&&&$ 0.380276(340)$&\\
        &3 &&&$-0.271613(040)$&&&$-0.271686(120)$&\\
        &4 &&&$0.325661(231) $&&&$ 0.324952(569)$&\\
        &5 &&&$-0.254040(074)$&&&$-0.254015(202)$&\\
        &6 &&&$0.312379(291) $&&&$ 0.311470(676)$&\\
        &7 &&&$-0.247721(103)$&&&$-0.247862(273)$&\\
        &8 &&&$0.307659(336) $&&&$ 0.306816(755)$&\\
        &9 &&&$-0.244968(132)$&&&$-0.245125(314)$&\\
        &10&&&$0.305604(361) $&&&$ 0.304557(803)$&\\
        &11&&&$-0.243631(158)$&&&$-0.243511(346)$&\\
        &12&&&$0.304598(379) $&&&$ 0.303233(784)$&\\
        &13&&&$-0.242977(179)$&&&$-0.242679(326)$&\\
        &14&&&$0.304154(390) $&&&$ 0.302762(822)$&\\
        &15&&&$-0.242729(188)$&&&$-0.242411(335)$&\\
        &16&&&$0.302465(998) $&&&$ 0.29796(295)$&\\
        \hline
    \end{tabular}
    \caption{The correlation function CF($r$) along $x$ axis of square lattice QDM under two different choices about $C$, $T=0.01$ $V=-0.5$ and $L=32$.}
    \label{Fig7}
\end{table}

\begin{figure}[htb]
    \includegraphics[clip,width=7cm]{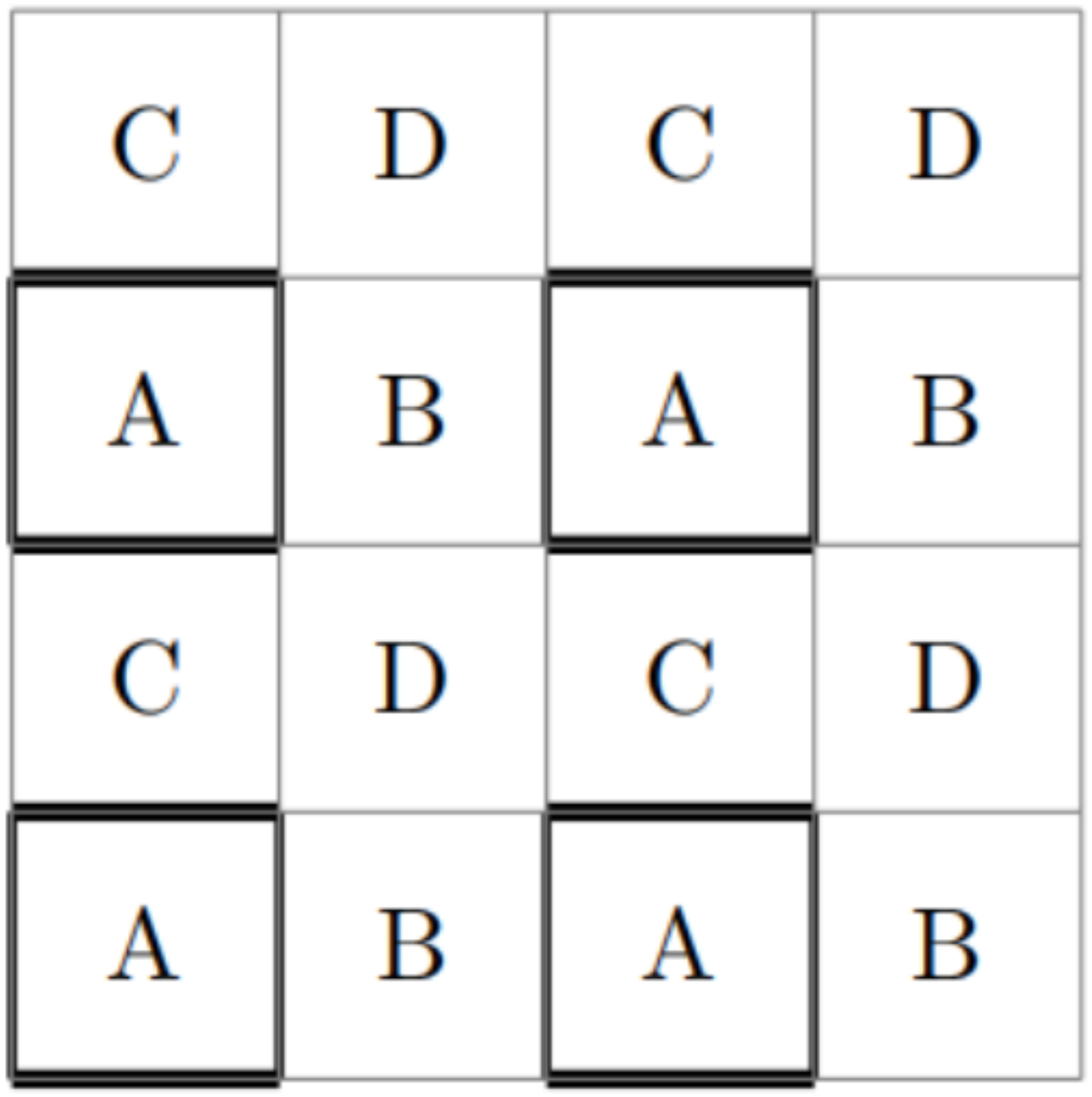}
    \caption{A, B, C and D four sublattices of mixed phase.}
    \label{Fig9}
\end{figure}

\begin{figure}[htb]
    \includegraphics[clip,width=8.6cm]{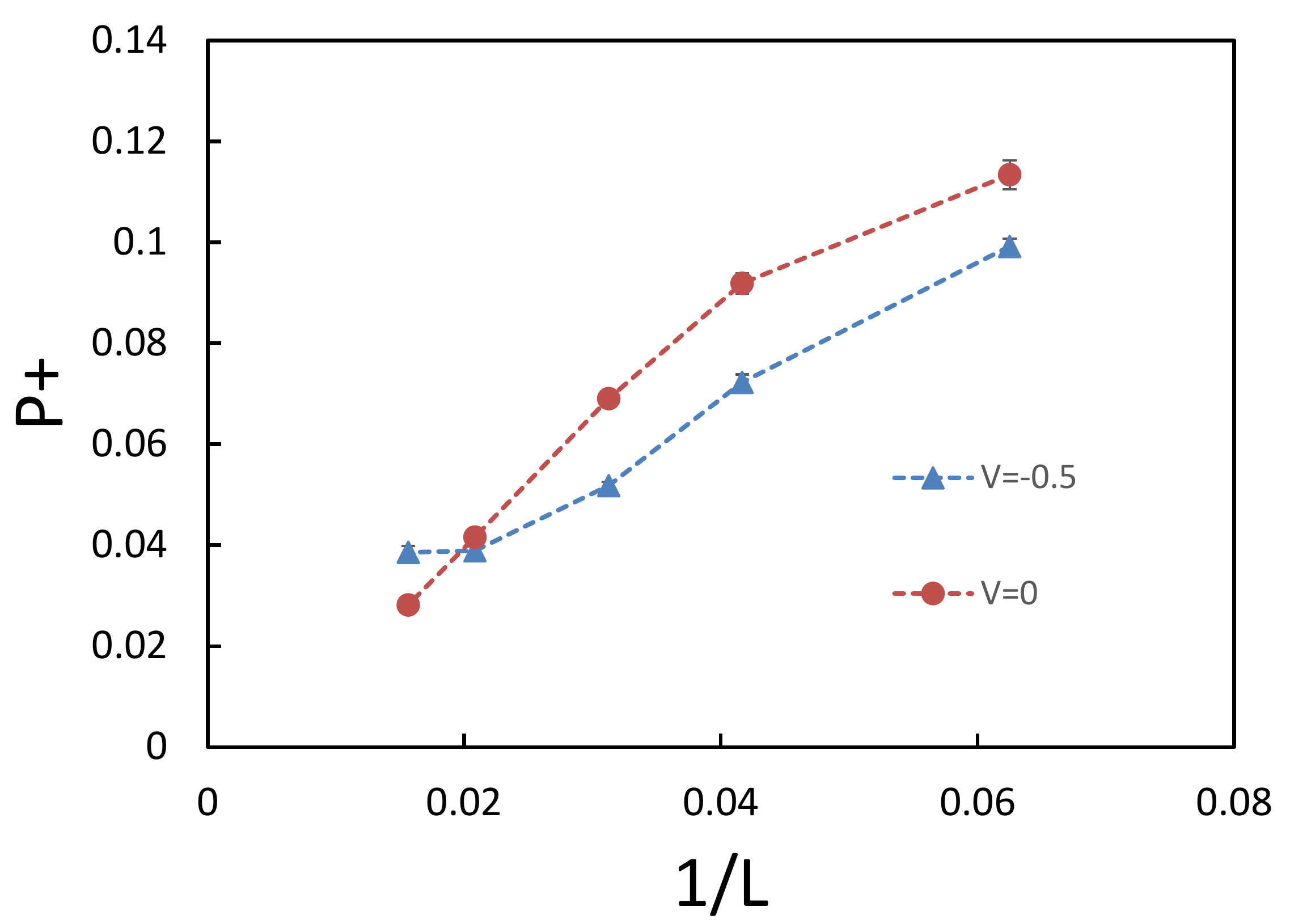}
    \caption{As size 1/L $\rightarrow$ 0, P+ tends to a finite value at V=-0.5 and 0 at V=0.}
    \label{Fig8}
\end{figure}

\subsection*{APPENDIX B: Sweeping cluster algorithm for $V<-1$ region}
In sweeping cluster method~\cite{ZY2019}, we write the QDM Hamiltonian in terms of plaquette operators $H_p$, $H=-\sum_{p=1}^{N_p}H_p$, where $p$ labels a specific plaquette on the lattice.  The plaquette operators are further decomposed into two operators: $H_p = H_{1,p} + H_{2,p}$, where $H_{1,p}$ is diagonal and $H_{2,p}$ is off-diagonal:
\begin{eqnarray}
H_{1,p} & = &
-V  \left( \vphantom{\sum} | \verdimers \rangle \langle \verdimers |+| \hordimers \rangle \langle \hordimers |\right) + V + C,
\label{hb1} \\
H_{2,p} & = & \left(  \vphantom{\sum} | \verdimers \rangle \langle \hordimers | + | \hordimers \rangle \langle \verdimers | \right).
\label{hb2}
\end{eqnarray}
We have subtracted a constant $N_p(V+C)$ from the QDM Hamiltonian, which should be kept in mind when calculating the energy. We do this because the constant $V+C$ makes all matrix elements of $H_{1,p}$ positive provided $C> {\rm max}(-V,0)$.

We choose $C=-V$ when $V<-1$ for simplicity. Although this will lose some of the updated vertices and make the update not ergodic enough, between the two alternative states~(columnar and mixed), this update can be fully migrating. The sweeping cluster algorithm works in region $(-\infty,0)$ when we choose $C=-V$ and works in $(-1,\infty)$ when $C=1$. The energies between $V=-1$ and 0 of different $C$ are same within error bar. Energy may be not good enough to verify the correctness of these two methods. We also compare a more sensitive physical quantity, that is, the correlation function as shown in Table~\ref{Fig7}. Here we choose $V=-0.5$ and $L=32$, CF($r$) means the correlation function~(CF) along $x$ axis, the distance is $r$. This also proves our choice of C when $V<-1$ is proper. In fact, the energies of these two methods are always similar on square lattice because ground states (no matter columnar or mixed phase) can be converted by pair update. However, on triangular lattice and other complex lattice, the energies will be significantly different due to the lack of ergodicity for the pair update~($C=-V$).\\

\subsection*{APPENDIX C: The original data of Fig.~4(a)}

We have shown all the original data of Fig.4(a) of article in Fig.~\ref{Fig6}(a) and (b) of appendix. All parameters are the same as defined in the Fig.3 (a) of article.

\subsection*{APPENDIX D: Another order parameters}

There was another order parameter $P_+$ to distinguish the mixed and columnar states as shown in Ref.\cite{mila2012}. In dimer language, $S_+(q)\sim \sum_{j,k}e^{iq(r_j-r_k)} [|\hordimers \rangle_j \langle \hordimers | | \hordimers \rangle_k \langle \hordimers |+| \hordimers \rangle_j \langle \verdimers | | \hordimers \rangle_k \langle \verdimers |] $. We have $P_+=S_+(\pi,\pi)^{1/2}$ to characterize the plaquette or mixed state in general.

$V=0$ is a special point not only due to this is a TFIM mapping point, but also because it is still in mixed phase though the $P_+$ falls down to 0. Though the period of mixed phase is (pi,pi), there is a particular case makes the $P_+=0$. As in the following Fig.\ref{Fig9}, if A=a, B=b, C=a-e, D=b-e, we could get a similar factor S(pi,pi)=A-B-C+D=0. However, it period is still 2 along x-axis and y-axis, i.e. (pi,pi). That's why $P_+=0$ but we can still measured non-zero $\Delta C_{\hordimers,\hordimers}$ and $\Delta C_{\hordimers,\verdimers}$.

At $V=-0.5$, we have a finite value even when $1/L\rightarrow0$, as shown in Fig.~\ref{Fig8}. It can be seen from Ref\cite{mila2012} that in the case of TFIM mapping (V=0 in QDM), the order parameter has already appeared an inflection point when $L$ is about $48$. Prove that the size effect begins to diminish at this scale. When V is smaller than 0, obviously the size effect should be weaker, that is, the corresponding order should be seen in smaller size. So this $V=-0.5$ result also prove the existence of mixed phase.

%

\end{widetext}

\end{document}